\documentclass[sigconf,authorversion,nonacm]{acmart}
\setcopyright{rightsretained}
\acmDOI{10.1145/3643774}
\acmYear{2026}
\copyrightyear{2026}
\acmJournal{PACMSE}
\acmVolume{1}
\acmNumber{ASE}
\usepackage{tcolorbox}
\usepackage[utf8]{inputenc}
\usepackage{array,multirow,etoolbox,graphicx,subcaption,fancyhdr,tabularx,longtable}
\usepackage{float}
\usepackage{graphicx}
\usepackage{amsmath} 
\usepackage{tcolorbox}
\usepackage{booktabs}
\usepackage{mathtools}
\usepackage{xcolor}
\usepackage{hyperref}
\newtcolorbox{TcolorBox}[1]{fonttitle=\bfseries,title=#1}
\usepackage{csquotes}
\usepackage[euler]{textgreek}
\usepackage{microtype}
\usepackage{tablefootnote}
\usepackage{threeparttable}
\newlength{\imagewidth}
\usepackage{amsmath,amsfonts}
\usepackage{algorithmic}
\usepackage{graphicx}
\usepackage{textcomp}
\usepackage{xcolor}
\usepackage{url}
\usepackage{microtype}
\usepackage{xurl}
\usepackage{tabularx}
\usepackage{caption}
\usepackage{subcaption}
\usepackage{xspace}
\usepackage{tikz}
\usetikzlibrary{shapes.geometric, arrows.meta, positioning, calc, fit, backgrounds}
\usepackage{enumitem}
\newlist{steps}{enumerate}{1}
\setlist[steps, 1]{label = Step \arabic*:}

\newcommand{\Meta}{Meta\xspace}

\newcommand{\rom}[1]{\uppercase\expandafter{\romannumeral #1\relax}}

\title{Automating Low-Risk Code Review at Meta: RADAR, Risk Calibration, and Review Efficiency}
\begin{document}
\begin{abstract}
\textbf{Context:} AI-assisted coding tools have altered software production. At \Meta, significant lines of code per human-landed diff grew by 105.9\% year over year and per-developer diff volume rose 51\%, with agentic AI responsible for over 80\% of that growth. Meanwhile, the share of diffs receiving timely review has declined, exposing a widening gap between code supply and reviewer bandwidth.
\textbf{Objective:} We ask three questions that progress from feasibility through calibration to impact: (1)~can risk-stratified automation operate at scale across diverse organizations, (2)~how does tuning the risk threshold affect the trade-off between automation yield and safety, and (3)~to what extent does automated review reduce end-to-end latency for AI-generated changes?
\textbf{Method:} We deployed RADAR (Risk Aware Diff Auto Review), a multi-stage funnel that classifies each diff by authorship and source type, applies eligibility gates, static heuristics, a machine-learned Diff Risk Score, LLM-based Automated Code Review, and deterministic validation before landing qualifying changes. We evaluate RADAR through telemetry covering 535K+ RADAR-reviewed diffs, observational before--after comparisons for policy changes, and difference-in-differences analysis of efficiency outcomes.
\textbf{Results:} RADAR has reviewed 535K+ diffs and landed 331K+. Relaxing the Diff Risk Score threshold from the 25th to the 50th percentile resulted in the approve rate to 60.31\%. The revert rate for RADAR-reviewed diffs is $\frac{1}{3}$ that of non-RADAR diffs, and the Production Incident (PI) rate is $\frac{1}{50}$ that of non-RADAR diffs. Compared to human-reviewed diffs, RADAR reduces median time to close by over 330\% and median diff review wall time by 35\%.
\textbf{Conclusion:} Risk-aware layered automation can materially reduce review bottlenecks created by AI-driven code growth without compromising production safety during incremental rollout.
\end{abstract}
\keywords{AI, Developer productivity}
\author{Chris	Adams,
Arjun Singh Banga,
Parveen	Bansal,
Souvik	Bhattacharya,
Payal Bhuptani,
Rujin Cao,
Pedro	Canahuati,
Nate	Cook,
Brian	Ellis,
Prabhakar	Goyal,
Gurinder	Grewal,
Tianyu He,
Matt 	Labunka,
Alex	Manners,
David	Molnar,
Ging Cee	Ng,
Vishal	Parekh,
Jiefu	Pei,
Frederic	Sagnes,
James 	Saindon,
Will	Shackleton,
Sid	Sidhu,
Gursharan	Singh,
Karthik Chengayan Sridhar,
Matt 	Steiner,
Pratibha 	Udmalpet,
Sean Xia,
Stacey	Yan,
Audris Mockus, 
Peter Rigby, 
Nachiappan Nagappan}
\orcid{}
\affiliation{
  \institution{Meta}
 \country{USA, UK, Canada}
}
\renewcommand{\shortauthors}{ }
\maketitle

\section{Introduction}\label{sec:intro}

Code review is a practice in modern software development. It improves correctness and maintainability.
Historically, at \Meta, reviewer capacity and tooling improvements have been sufficient to keep review latency within acceptable bounds. However, the process of code creation is changing, and the assumptions behind review at scale are being challenged.
Over the last year, significant lines of code per human landed diff increased by +105.9\% year over year, and diffs per developer per month increased by 51\% year over year, with 80\%+ of that increase attributed to agentic AI assistance. In parallel, the percentage of diffs reviewed within 24 hours is dropping. Together, these trends suggest that the rate of code creation is rising faster than human review capacity. In some large groups, we observed thousands of pending diff reviews, illustrating the problem of review backlogs. In this setting, a workflow that relies on humans reading most diffs does not scale, and we need mechanisms that preserve rigor while shifting scarce human attention toward changes where judgment and accountability are most beneficial.
To address this problem, we deployed RADAR, (Risk Aware Diff Auto Review), an end to end funnel that automates review and landing for a subset of low to medium complexity diffs while routing higher risk diffs to human review. RADAR combines eligibility checks, static heuristics, Diff Risk Scoring, LLM based automated code review, and deterministic validation checks. Rather than attempting to replace code review broadly, RADAR targets automation to diffs that satisfy conservative safety criteria and provides operational controls to support incremental rollout and tuning.
In this paper, we report production outcomes from the rollout of RADAR and organize our evaluation around three research questions that trace a logical progression from feasibility through calibration to developer impact:
\textit{RQ1 (Feasibility): Can risk-stratified automation operate at production scale, absorbing a meaningful share of review volume across diverse organizational contexts?}
\textit{RQ2 (Calibration): How does adjusting the risk threshold governing automation eligibility affect the trade-off between automation yield and safety outcomes?}
\textit{RQ3 (Impact): To what extent does automated review reduce end-to-end review latency for AI-generated code changes?}
Our results indicate that RADAR can operate at meaningful scale and is associated with large efficiency gains. To date, RADAR has reviewed 535K+ diffs and landed 331K+ diffs, reaching 25K+ diffs/day. The current RADAR-approve rate is 60.31\%, with a revert rate $\frac{1}{3}$ that of non-RADAR diffs and a PI rate $\frac{1}{50}$ that of non-RADAR diffs. Compared to human-reviewed diffs, RADAR reduces median time to close by over 330\% and median diff review wall time by 35\%.
This paper makes three contributions. First, we describe an end to end production workflow for risk aware automation of code review and landing (RACER), including distinct pipelines for automated sources and human authored diffs. Second, we present a differentiated eligibility model (Section~\ref{sec:diff_eligibility}) that applies distinct criteria to different automation source types, including deterministic codemods, AI-generated codemods, RACER runbooks with per-runbook risk history and volume controls, and human authored diffs with author and scope constraints. Third, we provide an initial empirical characterization of the rollout at scale, including throughput, calibration, and efficiency outcomes. These findings have implications for how organizations adapt review workflows under rapidly increasing diff volume and AI assisted code creation.
The remainder of this paper is structured as follows. Section~\ref{sec:background_motivation_process} provides background on review at \Meta, describes the RADAR process, and presents the diff eligibility model that determines which diffs qualify for automation based on authorship type, source classification, and organizational configuration. Section~\ref{sec:metrics} presents our study design and metric definitions. Section~\ref{sec:results} reports results for the three research questions. Section~\ref{sec:threats} discusses threats to validity and we conclude with implications for future review automation.

\section{Background, Motivation, and RADAR Process}
\label{sec:background_motivation_process}

\subsection{Software Development at \Meta}
\label{s:sdmeta}

\Meta develops software for both server side services and client facing products, including specialized hardware devices. This approach supports frequent software updates and provides strong control over versioning and configurations. At \Meta, this deployment strategy has cultivated a routine of frequently pushing new code to production. Before a change is deployed, it typically undergoes peer review, automated testing, and staged rollout practices, including canary and incremental deployments. After deployment, engineers monitor logs and other telemetry to detect potential issues and to support rapid rollback when necessary.
Code review is a standard part of software development at \Meta and serves multiple functions \cite{Shan2022FSE}. First, it encourages authors to maintain high coding standards because changes will be examined by peers. Second, reviewers can catch defects, improve readability, and suggest alternative designs.
These benefits are important in a large organization and become more important as the codebase evolves and as engineers change projects.
\Meta uses Phabricator\footnote{\url{http://phabricator.org}} to support contemporary code review and continuous integration. Developers submit code changes that are referred to as \textit{diffs} at \Meta, describe the change and test plan, and then request review from one or more reviewers or reviewer groups. Reviewers provide feedback through inline comments and actions, and authors iterate through additional revisions until the diff is accepted and incorporated into the codebase, which is called landing, or until the diff is abandoned.
Phabricator and version control systems together provide the operational data needed to study review workflows. Phabricator tracks diff metadata, author and reviewer actions, timestamps for major lifecycle events, and the current state of each diff. A diff consists of one or more patches representing the initial version of a change and subsequent revisions made in response to review feedback. This paper focuses on review at \Meta under rapidly increasing change volume, and on workflow support that combines automation with conservative policy and validation to help scale review without removing accountability.
\subsection{RACER}\label{sec:racer}
RACER, Risk-Aware Code Editing and Refactoring, is \Meta's GenAI-powered tool for automating code changes at scale. RACER enables developers to delegate well-defined coding tasks, such as dead code removal, reducing cyclomatic complexity, framework migrations, lint fixes, and test generation, to an AI agent that generates diffs for human review.
RACER operates through \textit{runbooks}, which are pre-configured prompts that encode the instructions, constraints, and context for a specific type of code change. When a task is assigned, RACER picks it up, generates the code change in a sandboxed environment, runs verification steps, and submits the resulting diff. Developers review the diff and can provide feedback through inline comments; RACER iterates on the diff in response to feedback, retrying on build, test, or static analysis errors. At the time of this study, RACER was landing approximately 3,000 diffs per week, with 59\% of changes landing without human revisions.
RACER also supports autonomous codebase maintenance through \textit{RACER Maintainer}, which manages the full lifecycle of bulk code changes: discovering issues from code quality dashboards, creating tasks, generating diffs, assigning reviewers, and managing the diff queue. This mode is used for high-volume use cases such as cyclomatic complexity reduction, quality insight remediation, broken test fixes, and framework migrations.
RACER is relevant to RADAR because RACER-generated diffs are one of the primary sources of bot-authored diffs that enter the RADAR pipeline. When RACER runbooks are onboarded to RADAR, their diffs can be RADAR-reviewed and RADAR-landed through the ACE policy (Section~\ref{sec:radar_ai_pipeline}) without requiring any human review, provided they pass all safety checks.
\subsection{Diff Risk Score}\label{sec:drs}
Diff Risk Score (DRS) is a machine learning model that predicts how likely a diff is to cause a negative outcome, primarily a production incident (PI). DRS assigns a continuous risk score to every diff and is optimized for high recall at a given percentage of diffs flagged: for example, flagging 10\% of diffs while catching 60\% of PI-causing changes. The score is expressed as a percentile (PX), where PX means that only the lowest-risk X\% of diffs qualify for a given threshold. A lower threshold is more conservative: P5 means only the safest 5\% of diffs pass, while P50 means the safest 50\% pass.
DRS was originally developed as a signal for the OrgA to allow low-risk diffs to land during code freezes. It has since evolved into a broader risk platform that powers approximately 20 risk-aware features across \Meta, including accept-to-ship workflows, cherry-pick risk assessment for mobile releases, code refactoring prioritization, and reviewer recommendations for high-risk diffs.
In RADAR, DRS serves as a gating mechanism at multiple stages of the pipeline. For AI and bot diffs, DRS thresholds determine whether a diff is eligible for RADAR-landing: allowlisted RACER runbooks use a P50 threshold, while non-allowlisted sources use a stricter P20 threshold (Section~\ref{sec:radar_ai_pipeline}). For human-authored diffs, the default DRS threshold is P5, meaning only the lowest-risk 5\% of diffs qualify; RADAR Approval applies additional constraints to waive deferred review entirely (Section~\ref{sec:radar_human_pipeline}). Each organization can configure its own DRS thresholds based on its risk appetite.
\subsection{Automated Code Review}\label{sec:acr}
Automated Code Review (ACR) is an LLM-based code review agent that reads and interprets the actual code changes in a diff, going beyond metadata and static heuristics to provide semantic understanding of what was changed. ACR is integrated into \Meta's code review workflow and serves two roles: providing comprehensive review feedback (general and inline comments) for diff authors and reviewers, and making automated accept or reject decisions for the RADAR pipeline.
The ACR evaluation pipeline classifies each change in a diff against a set of predefined safe and risk signals. Safe signals include refactoring without behavioral change, dead code removal, defensive programming additions, logging additions, pure formatting changes, documentation and comment updates, import hygiene, test additions, and static resource updates. Risk signals include high calculated review effort (complexity score of 4 or above), substantial structural changes, identified bugs or logic errors, performance risks, and security vulnerabilities such as secrets exposure, SQL injection, or authentication bypasses.
For a diff to be auto-accepted, ACR requires a confidence score of at least 8 out of 10, with all changes classified into safe categories. If any risk signal is detected, the diff is automatically disqualified from auto-acceptance. This conservative acceptance criterion ensures that only changes with high confidence of safety proceed through the automated pipeline.
ACR adds a critical layer of semantic understanding that static heuristics cannot provide. Static checks operate on metadata, file paths, and thresholds, but they cannot determine what a code change actually does. ACR enables RADAR to approve changes that static checks might flag as risky but are actually safe, such as large but trivial refactors or mechanical dead code cleanup, and to catch subtle risks that pattern matching cannot detect, such as logic errors or unintended behavioral changes. Together with its successor DCR, ACR forms the \textit{RADAR Review Agent}, the AI-powered code review engine used in both the ACE pipeline for bot diffs (Section~\ref{sec:radar_ai_pipeline}) and the RADAR Verification pipeline for human-authored diffs (Section~\ref{sec:radar_human_pipeline}).
\subsection{Motivation: Review at Scale}\label{sec:scale_background}
\begin{figure*}[t]

\centering
\includegraphics[width=\textwidth]{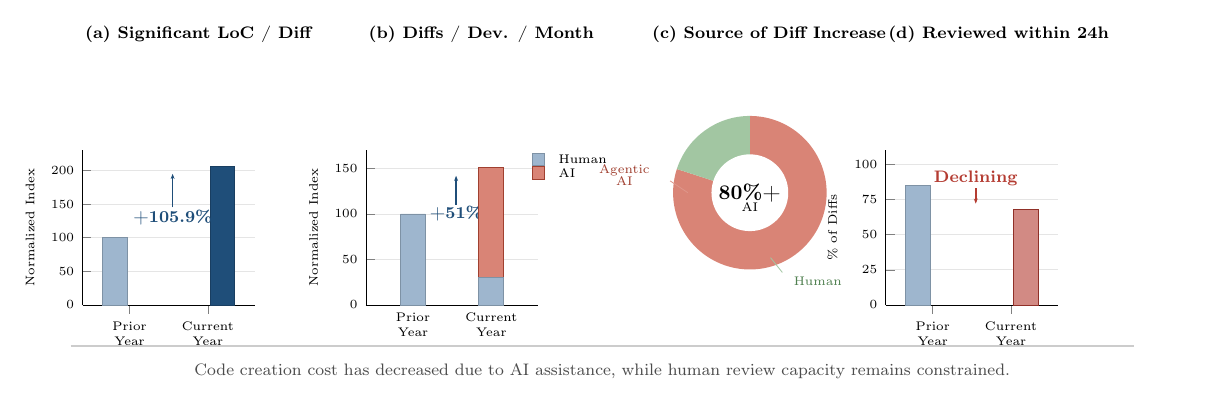}

\caption{Scale pressure on code review at \Meta. Over the last year, significant lines of code per human landed diff increased by +105.9\% year over year, and diffs per developer per month increased by 51\% year over year, with 80\%+ of that increase attributed to agentic AI assistance. In the same period, the percentage of diffs reviewed within 24 hours dropped, indicating that review capacity is not keeping pace with code creation.}
\label{fig:scale_problem}

\end{figure*}

Code review is an important activity in the software development lifecycle that improves correctness, reliability, and shared understanding. At \Meta, it is also a primary coordination mechanism, as engineers rely on review to
align changes across a large monorepo.
However, the nature of code creation have shifted. Over the last year, significant lines of code per human landed diff increased by +105.9\% year over year, and diffs per developer per month increased by 51\% year over year, with 80\%+ of that increase attributed to agentic AI assistance. In parallel, the percentage of diffs reviewed within 24 hours is dropping. These trends indicate that code creation has accelerated, while review remains constrained by human attention and calendar time.
To help the reader interpret these shifts, Figure~\ref{fig:scale_problem} summarizes the year over year trends in diff size, diff volume, and review timeliness that motivate this work.
The scaling pressure is visible in product areas with high diff volume. In this context, our goal is to preserve rigor while shifting human attention toward changes that require human judgment, and toward diffs that are more likely to introduce risk. This motivates a process where routine low risk diffs can proceed with minimal human involvement, and higher risk diffs are routed to appropriate reviewers with clear ownership.

\subsection{RADAR Overview}\label{sec:radar_overview}

RADAR, Risk Aware Diff Auto Review, is our approach to scaling review capacity while maintaining safety. RADAR is designed as an end to end funnel rather than a single model. The funnel combines static heuristics, Diff Risk Scoring, and an LLM based automated review agent, and it uses conservative policy gates and operational controls to manage risk.
RADAR is designed to support incremental rollouts. We gate automation by authorship type and eligibility, we enforce onboarding requirements for automated sources, and we include operational controls that allow us to pause automation for use cases that trigger safety incidents. These controls allow us to expand coverage gradually, monitor outcomes, and adjust thresholds without changing the user facing review tool.

\subsection{Diff Eligibility}\label{sec:diff_eligibility}

\begin{figure*}[t]

\centering
\includegraphics[width=\textwidth]{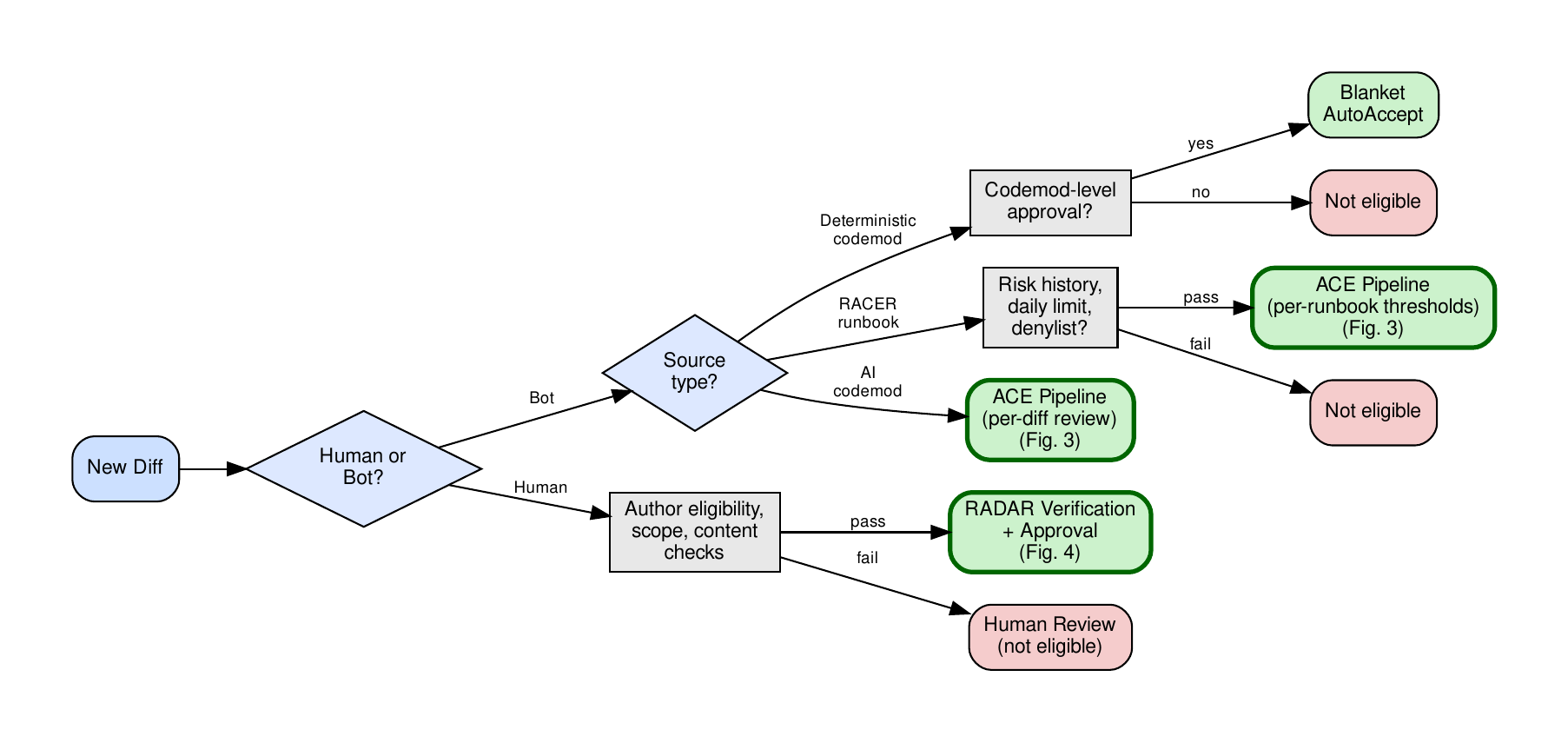}

\caption{Diff eligibility classification in RADAR. Each diff is first classified by authorship type (human or bot), then bot diffs are further classified by automation source. Deterministic codemods can bypass per-diff AI review through Blanket AutoAccept. AI-generated codemods require per-diff ACE evaluation. RACER runbooks must pass per-runbook eligibility checks (risk history, daily limits, denylist) before entering the ACE pipeline with runbook-specific DRS thresholds. Human diffs must satisfy author eligibility, scope, and content constraints before entering the RADAR Verification and Approval pipeline.}
\label{fig:eligibility_classification}

\end{figure*}

The important design decision in RADAR is how diffs are classified and whether they are eligible for automated review and landing. Eligibility is not a single check but a layered evaluation that depends on the authorship type of the diff, the specific automation source, and the organizational context. This subsection describes the eligibility model as a contribution distinct from the pipeline mechanics, because the eligibility determination itself encodes the risk management strategy and explains the scope of automation.
Figure~\ref{fig:eligibility_classification} illustrates the eligibility classification tree that routes each diff to the appropriate review path based on authorship type and automation source.
\subsubsection{Authorship Classification}
RADAR first classifies each diff by authorship type: human authored or bot authored. Bot authored diffs are further classified by their automation source, which determines the specific eligibility path. This classification is important because human and bot diffs differ in risk profile and volume characteristics.
Human authored diffs enter the RADAR Verification and Approval pipeline (Section~\ref{sec:radar_human_pipeline}), where eligibility depends on author attributes and diff metadata. Bot authored diffs enter the ACE pipeline (Section~\ref{sec:radar_ai_pipeline}), where eligibility depends on the automation source, its track record, and per-source configuration.
\subsubsection{Bot Diff Eligibility by Source Type}
Not all bot diffs are treated equally. RADAR distinguishes three categories of automation sources, each with a distinct eligibility model:
\paragraph{Deterministic codemods (Blanket AutoAccept).} CodemodService supports deterministic code transformations such as API migrations, import reorganization, and mechanical refactoring. When a codemod is classified as deterministic, meaning the transformation is fully specified and does not involve LLM generation, it can be approved through a Blanket AutoAccept policy. Diffs from these codemods bypass per-diff AI review entirely because the transformation itself has been vetted at the codemod level rather than at the diff level. This is the least restrictive eligibility path, and it requires that the codemod has been reviewed and approved for blanket auto-acceptance.
\paragraph{AI-generated codemods (Conditional AutoAccept).} When a CodemodService configuration uses LLM-based generation, meaning each diff may vary depending on the code context and model output, the diffs cannot be blanket-approved. Instead, each diff must individually pass through the ACE pipeline, including Diff Risk Scoring and Automated Code Review. This Conditional AutoAccept path ensures that the variability inherent in AI-generated changes receives per-diff scrutiny.
\paragraph{RACER runbooks.} RACER-generated diffs (Section~\ref{sec:racer}) follow the most granular eligibility model. Each RACER runbook is independently evaluated for eligibility based on four criteria:
\begin{enumerate}[leftmargin=*]
\item \textbf{Risk history heuristics.} Each runbook must demonstrate a clean track record over a 60-day lookback window: zero PIs, a low revert rate, a low human rejection rate, and a minimum number of landed diffs to establish statistical confidence.
\item \textbf{Per-runbook daily limits.} To prevent any single runbook from flooding the commit queue, daily RADAR-landing caps are enforced. Default limits are conservative, but runbooks with established safety records can be elevated to higher limits (up to 2,000 diffs per day for high-volume use cases).
\item \textbf{Per-runbook DRS thresholds.} Diff Risk Score thresholds are configured per runbook. Allowlisted runbooks with strong safety records use a relaxed P50 threshold, while non-allowlisted runbooks use the stricter default P20 threshold.
\item \textbf{Denylist.} Specific runbooks that have caused incidents or that target sensitive areas are permanently blocked from RADAR-landing. Additionally, runbooks whose names contain certain keywords (e.g., ``test'') are excluded to avoid automating changes to test infrastructure without human oversight.
\end{enumerate}
This per-runbook granularity is a distinctive feature of RADAR's eligibility model: two RACER runbooks may have identical code transformations, but if one has a history of reverts, it will be blocked while the other proceeds.
\subsubsection{Human Diff Eligibility}
Human authored diffs are eligible for automated review through RADAR Verification (Section~\ref{sec:radar_human_pipeline}) when they satisfy author, scope, and content constraints:
\begin{enumerate}[leftmargin=*]
\item \textbf{Author eligibility.} The author must satisfy internal eligibility criteria, including role/experience and operational ownership requirements.
These criteria ensure that automated review is applied to diffs from engineers who have established context and accountability.
\item \textbf{Scope exclusions.} Diffs that touch open-source code, SOX-scoped code, or code requiring additional reviews are excluded. These exclusions protect areas where constraints require human judgment.
\item \textbf{Diff state constraints.} The diff must not be work-in-progress, a request for comments, or previously rejected. It must be the latest published version, and CI signals must be in an allowed state.
\item \textbf{Content-level checks.} The diff must not contain blocklisted code phrases, and must not touch files matching suffix or prefix blocklists for certain configurations and paths.
\end{enumerate}
\subsubsection{Organizational Configuration}

Eligibility thresholds are not fixed globally. Each organization can configure its own risk appetite through \textit{OrgRADARPolicyConfig}, which controls DRS thresholds, whether deferred review is enabled, and which automation sources are permitted. For example, OrgA uses a relaxed bot DRS threshold (effectively bypassing DRS gating for bot diffs and relying on ACR directly), while other organizations maintain stricter defaults managed by the OrgB team. This organizational configurability allows RADAR to operate across diverse risk environments within the same company.
Table~\ref{tab:eligibility_summary} summarizes the eligibility criteria by diff source type.
\begin{table*}[t]

\centering
\caption{Summary of RADAR diff eligibility criteria by source type. Each source type has distinct eligibility requirements that reflect its risk profile and operational characteristics.}

\small
\setlength{\tabcolsep}{4pt}
\renewcommand{\arraystretch}{1.2}
\begin{tabular}{ p{.18\imagewidth} | p{.22\imagewidth} | p{.18\imagewidth} | p{.18\imagewidth} | p{.18\imagewidth} }
{\bf Criterion} & {\bf Deterministic Codemod} & {\bf AI Codemod} & {\bf RACER Runbook} & {\bf Human Author} \\ \hline
Pipeline & Blanket AutoAccept & ACE & ACE & RADAR Verification + Approval \\ \hline
Per-diff AI review & No & Yes (ACR) & Yes (ACR) & Yes (ACR) \\ \hline
DRS gating & No & Yes & Yes & Yes (per-org configurable) \\ \hline
Risk history required & No (codemod-level review) & No & Yes (60-day lookback) & No \\ \hline
Daily volume limits & No & No & Yes (per runbook) & No \\ \hline
Author/source vetting & Codemod approval & Codemod config review & Runbook onboarding & Eligible role, oncall, tenure \\ \hline
Scope exclusions & Standard & Standard & Standard + denylist & Standard + file blocklists \\
\end{tabular}
\label{tab:eligibility_summary}

\end{table*}

\subsection{RADAR Pipeline for AI and Bot Diffs}\label{sec:radar_ai_pipeline}

\begin{figure*}[t]

\centering
\includegraphics[width=\textwidth]{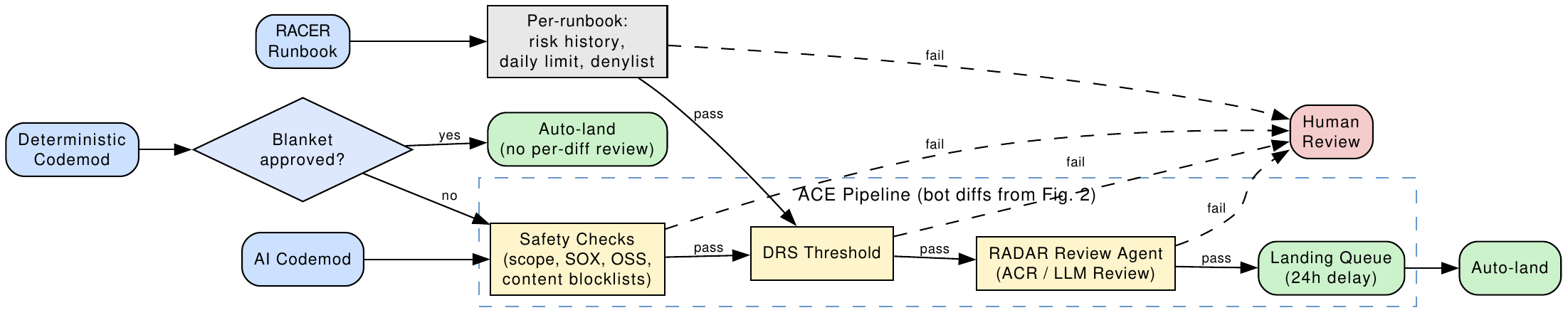}

\caption{RADAR pipeline for AI and bot diffs. Different bot source types follow distinct paths. Deterministic codemods with Blanket AutoAccept bypass the ACE pipeline entirely. AI-generated codemods enter the ACE pipeline directly. RACER runbooks must first pass per-runbook eligibility checks (risk history, daily limits, denylist) before entering ACE with runbook-specific DRS thresholds.
Within ACE, all diffs pass through three layers: safety checks, Diff Risk Score, and the RADAR Review Agent (LLM-based code review). Approved diffs land after a configurable delay.
Failure at any stage routes the diff to human review.}
\label{fig:radar_ai_process}

\end{figure*}

For bot diffs that pass the eligibility checks described in Section~\ref{sec:diff_eligibility}, RADAR uses a policy called ACE, AI Commit Eligibility. ACE is intended to allow changes from automated sources to land without any human review when they satisfy strict safety criteria. As shown in Figure~\ref{fig:radar_ai_process}, different bot source types follow distinct paths through the pipeline: deterministic codemods with Blanket AutoAccept bypass ACE entirely, while AI-generated codemods and eligible RACER runbooks enter the ACE pipeline with source-specific DRS thresholds. The ACE pipeline evaluates each entering diff using three validation layers that must all pass: static heuristics, Diff Risk Scoring thresholds, and Automated Code Review.
Figure~\ref{fig:radar_ai_process} shows the ACE pipeline, including the differentiated paths for each bot source type and the outcomes of auto land versus routing to human review.
The ACE pipeline evaluates each bot diff through three validation layers, all of which must pass. The first layer applies static heuristics that filter out diffs violating hard constraints: the diff must not be open source, must not require additional reviews, must not touch SOX-scoped code, and must originate from an onboarded automation source, either a RACER runbook (Section~\ref{sec:racer}) or a vetted CodemodService configuration. Only automation sources with an established track record of safe landings, low revert rates, and non-controversial changes are eligible for onboarding.
The second layer evaluates the Diff Risk Score (Section~\ref{sec:drs}), comparing it against configurable thresholds based on organizational risk appetite. For OrgA and partner organizations, allowlisted runbooks with established safety records are assigned a P50 threshold, meaning that the safest 50\% of diffs qualify. Non-allowlisted runbooks and other AI bots are subject to a stricter P20 threshold. In other organizations, default DRS thresholds are maintained by the OrgB team.
The third layer is the RADAR Review Agent (Section~\ref{sec:acr}), which uses an LLM to read the actual code changes and classify them against safe and risk signals. The review agent verifies that no business logic requiring human judgment was updated, and detects safe non-functional patterns such as dead code removal, refactoring, error handling improvements, logging additions, and test updates. For functional changes, it evaluates whether the changes are correct and low-risk enough to not require human review. If any risk signal is detected, the diff is rejected and routed to human review.
Because AI and bot diffs can be produced at high volume, the ACE pipeline also includes operational limits to prevent flooding, including limits on auto landing windows and caps on diffs per runbook per day, such as limiting RACER runbooks to 10 to 2000 diffs generated per day. Approved bot diffs land after a configurable delay, during which a human reviewer can still reject the diff. We also implement feedback loops to monitor the funnel, identify blockers, and tune thresholds.
\subsection{RADAR Pipeline for Human Authored Diffs}
\label{sec:radar_human_pipeline}
\begin{figure*}
\centering
\includegraphics[width=\textwidth]{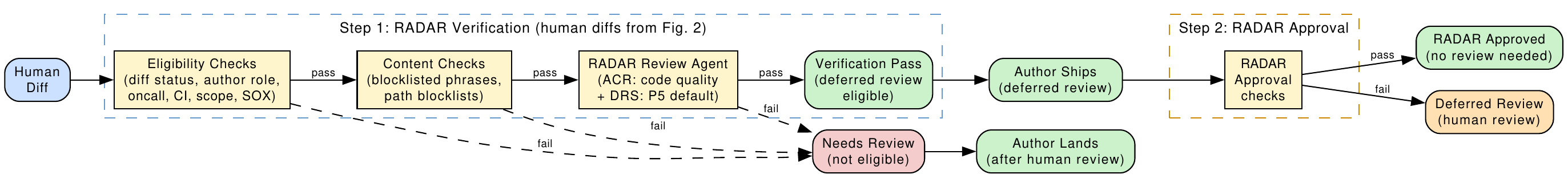}
\caption{RADAR pipeline for human authored diffs. Step~1 (RADAR Verification) evaluates diffs through eligibility checks, content checks, and the RADAR Review Agent with DRS thresholds (P5 default) to determine whether a diff can ship with deferred human review. Step~2 (RADAR Approval) applies stricter criteria to verified diffs; those that pass require no human review at all. Failure at any verification stage routes the diff to standard human review.}
\label{fig:radar_human_process}
\end{figure*}

For diffs authored by humans, RADAR uses a two-step process: \textit{RADAR Verification} followed by \textit{RADAR Approval}. RADAR Verification
determines whether a human-authored diff can safely land with a deferred post-land human review. RADAR Approval then evaluates verified diffs against stricter criteria to determine whether the deferred review can be waived entirely, meaning no human review is required. Both steps combine static heuristics, Diff Risk Score (Section~\ref{sec:drs}), and the RADAR Review Agent (Section~\ref{sec:acr}) in a layered funnel where all checks must pass. The author always retains full control: they can ship with RADAR Approval, wait for a human reviewer, or return the diff to ``Needs Review'' status.
RADAR Verification operates in three sequential groups. The first group applies eligibility checks based on metadata and safety constraints. Diff status and marker checks require that the diff must not be work-in-progress or a request for comments, must not have been rejected, must not be in code freeze, and must be the latest published version. Author eligibility requires an eligible role (SWE job family including management, Data Engineers, Data Scientists, or anyone who has committed more than 10 diffs in the past year), with SWE interns requiring at least 60 days of employment, and an associated operational oncall rotation. Scope exclusions filter out open-source diffs, SOX-scoped code, and diffs requiring additional review. Additionally, Continuous Integration (CI) signals must be in an allowed state. The second group applies content-level checks: diff content must not contain blocklisted code phrases, and the diff must not touch files matching suffix or prefix blocklists. The third group invokes the RADAR Review Agent for LLM-based semantic analysis of the code changes, combined with DRS evaluation against a configurable threshold (P5 default, meaning only the lowest-risk 5\% of diffs qualify). If the diff passes all three groups, the author is notified that the diff is verified and can ship immediately with a deferred review.
Verified diffs are then automatically evaluated by RADAR Approval against stricter criteria. Diffs that pass receive RADAR Approval, meaning no human review is required---not even deferred. Figure~\ref{fig:radar_human_process} shows the two-step pipeline from RADAR Verification through RADAR Approval.
RADAR Verification checks metadata such as author eligibility, safety constraints, and diff status. It then applies content-level checks and invokes the RADAR Review Agent for semantic code analysis, combined with DRS evaluation against the configured threshold. After verification, the author can ship the diff with deferred review. RADAR Approval then evaluates the verified diff against stricter criteria to determine whether the deferred review can be waived entirely, so that no human review is required.
This section provides the context and the process used in the remainder of the paper. In the following sections, we describe our evaluation methodology and outcome metrics, and we report on controlled experiments that measure both goal outcomes and guardrail outcomes of RADAR interventions.
\section{Study Design}
\label{sec:metrics}
In this paper, we evaluate RADAR, Risk Aware Diff Auto Review, an end to end funnel that combines eligibility checks, static heuristics, Diff Risk Scoring, LLM based automated code review, and deterministic validation checks. Our goal is to characterize how RADAR performs 
as it scales, and to quantify observed associations with efficiency outcomes while reporting safety related outcomes that are monitored during rollout.
The evidence available for RADAR in this paper is based on production rollouts, descriptive operational measurement at launch scale, observational comparisons across time, and difference in differences style comparisons over large datasets. We structure our evaluation around three research questions. Each research question is answered using the measurement definitions described in Section~\ref{sec:metrics}. 
we report time windows at the level of weeks and months.
\paragraph{RQ1 (Feasibility): Automated handling at scale.}
\textit{Can risk-stratified automation operate at production scale, absorbing a meaningful share of review volume across diverse organizational contexts?}
For RQ1, we report descriptive operational measurements from the production rollout. The unit of analysis is the diff. We report aggregate counts for RADAR reviewed diffs and RADAR landed diffs, peak daily throughput for RADAR reviewed diffs, the number of organizations covered, and safety outcomes including reverts and PIs.
\paragraph{RQ2 (Calibration): Risk threshold tuning.}
\textit{How does adjusting the risk threshold governing automation eligibility affect the trade-off between automation yield and safety outcomes?}
For RQ2, we evaluate a production policy change where the Diff Risk Score threshold was relaxed from p25 to p50. This is an observational comparison across time, where the unit of analysis is the diff. We report changes in automation yield, measured by RADAR approve rate, and changes in deterministic validation outcomes, measured by RADAR Verification pass rate
We also report safety related outcomes that were monitored during the same period, including PIs and reverts.
\paragraph{RQ3 (Impact): Review efficiency for AI-generated diffs.}
\textit{To what extent does automated review reduce end-to-end review latency for AI-generated code changes?}
For RQ3, we use a large scale, difference in differences style comparison to estimate changes in time to close and diff review wall time for AI generated diffs as RADAR coverage increases. The unit of analysis is the diff. Because the analysis is not based on random assignment and we do not provide the full specification of the underlying model, we report the measured deltas as observational estimates.
\paragraph{Metrics.}
Across all research questions, we report throughput and coverage measures, automation yield and validation measures, efficiency measures, and safety related measures that were monitored during rollout. Table~\ref{tab:RADARStudyOverview} summarizes the study design for each research question and the reporting windows. Table~\ref{tab:RADARStudyMetrics} summarizes the metrics used in each research question.
\begin{table*}[t]
\centering
\caption{Overview of the study design used to evaluate RADAR. The evidence includes descriptive operational measurement, observational comparisons across time, and difference in differences style comparisons over large datasets.}
\small
\setlength{\tabcolsep}{4pt}
\renewcommand{\arraystretch}{1.2}
\begin{tabular}{ p{.16\imagewidth} | p{.36\imagewidth} | p{.18\imagewidth} | p{.12\imagewidth} | p{.18\imagewidth} }
{\bf RQ} & {\bf Description} & {\bf Method} & {\bf Unit} & {\bf Time period / scale} \\ \hline
RQ1 & Throughput and coverage during rollout & Descriptive operational measurement & Diff & 535K+ RADAR reviewed diffs, 331K+ RADAR landed diffs. \\ \hline
RQ2 & Policy change, p25 to p50 & Observational comparison across time & Diff & Threshold changed from p25 to p50, 535K+ RADAR reviewed diffs \\ \hline
RQ3 & Efficiency for AI generated diffs & Difference in differences style comparison over large datasets & Diff & 535K+ diffs reviewed by RADAR  \\
\end{tabular}
\label{tab:RADARStudyOverview}
\end{table*}
\begin{table*}[t]
\centering
\caption{Metrics used to evaluate RADAR by research question.}
\small
\setlength{\tabcolsep}{4pt}
\renewcommand{\arraystretch}{1.2}
\begin{tabular}{ p{.30\imagewidth} | p{.48\imagewidth} | p{.07\imagewidth} | p{.07\imagewidth} | p{.08\imagewidth} }
{\bf Metric} & {\bf Description} & {\bf RQ1} & {\bf RQ2} & {\bf RQ3} \\ \hline
RADAR reviewed diffs & Number of diffs processed by RADAR & Yes & NA & NA \\ \hline
RADAR landed diffs & Number of diffs landed by RADAR & Yes & NA & NA \\ \hline
Daily throughput & Peak daily diffs RADAR reviewed & Yes & NA & NA \\ \hline
Coverage & Number of organizations covered & Yes & NA & NA \\ \hline
RADAR approve rate & Fraction of eligible diffs that are RADAR approved & NA & Yes & NA \\ \hline
Verification pass rate & Fraction of diffs passing RADAR Verification & NA & Yes & NA \\ \hline
Time to close & End to end time from diff publish to diff close (median) & NA & NA & Yes \\ \hline
Diff review wall time & Wall time diffs spend waiting for and undergoing review (median) & NA & NA & Yes \\ \hline
Reverts & Count of landed diffs that are reverted & Yes & Yes & NA \\ \hline
PIs & Count of safety incidents attributable to landed diffs & Yes & Yes & NA \\
\end{tabular}
\label{tab:RADARStudyMetrics}
\end{table*}
\section{Results}
\label{sec:results}
\subsection{RQ1 (Feasibility): Automated Handling at Scale}\label{sec:rq1_automated_handling}
\textit{Can risk-stratified automation operate at production scale, absorbing a meaningful share of review volume across diverse organizational contexts?}
\begin{table*}[t]
\centering
\caption{RQ1. Descriptive operational outcomes for RADAR.}
\begin{tabularx}{.85\textwidth}{X | r}
\textbf{Metric Name} & \textbf{Reported Value} \\ \hline
RADAR reviewed diffs & 535,290 \\ \hline
RADAR landed diffs & 331,720 \\ \hline
Reverts (out of RADAR reviewed diffs) & $\frac{1}{3}$ the rate of non-RADAR diffs (p-val$<1e^{16}$ Fisher's exact test)\\ \hline
PIs (out of RADAR reviewed diffs) & $\frac{1}{50}$ the rate of non-RADAR diffs\footnotemark (p-val$<1e^6$ Fisher's exact test) \\
\end{tabularx}
\label{tab:rq1_results}

\end{table*}
\footnotetext{All PIs were manually reviewed by domain experts and none were deemed to have been detected by a human reviewer}
RADAR was designed to address a review capacity bottleneck, where review capacity is constrained by human attention, while code creation has accelerated. Over the last year, significant lines of code per human landed diff increased by +105.9\% year over year and diffs per developer per month increased by 51\% year over year, with 80\%+ of that increase attributed to agentic AI assistance. The growing review queue motivated an approach that shifts human attention toward higher risk changes and uses automation for eligible low risk diffs.
Table~\ref{tab:rq1_results} shows that RADAR has reached substantial scale, reviewing 535K+ diffs and landing 331K+ diffs. Daily throughput reached 25K diffs RADAR-reviewed. The revert rate is $\frac{1}{3}$ that of non-RADAR diffs, and the PI rate is $\frac{1}{50}$ that of non-RADAR diffs. These scale and safety outcomes indicate that an end to end funnel that combines eligibility checks, Diff Risk Scoring, LLM based automated code review, and deterministic validation can operate at a volume that is relevant to the review backlog problem while maintaining strong safety outcomes.
An important implication is that automation can absorb a portion of routine diff volume without requiring a proportional increase in human reviewer capacity. Given that the percentage of diffs reviewed within 24 hours is dropping, throughput improvements of this magnitude can matter even if the automation scope is limited to low and medium complexity diffs.
\begin{tcolorbox}
RADAR has reviewed and landed thousands of diffs, with daily throughput reaching 25K diffs reviewed. The revert rate is $\frac{1}{3}$ that of non-RADAR diffs, and the PI rate is $\frac{1}{50}$ that of non-RADAR diffs.
\end{tcolorbox}
\subsection{RQ2 (Calibration): Risk Threshold Tuning}
\label{sec:rq2_drs_calibration}
\textit{How does adjusting the risk threshold governing automation eligibility affect the trade-off between automation yield and safety outcomes?}
\begin{table*}[t]

\centering
\caption{RQ2. Observational outcomes associated with relaxing the Diff Risk Score threshold from p25 to p50. We report aggregate values and do not report statistical tests.}

\begin{tabularx}{.85\textwidth}{X | r}
\textbf{Metric Name} & \textbf{Reported Value} \\ \hline
DRS threshold policy change & p25 to p50 \\ \hline
RADAR approve rate (L7, current) & 60.31\% \\ \hline
RADAR Verification pass rate (L7, current) & 26.31\% \\ \hline
Reverts across 535K+ RADAR reviewed diffs & $\frac{1}{3}$ the rate of non-RADAR diffs \\ \hline
PIs across 535K+ RADAR reviewed diffs & $\frac{1}{50}$ the rate of non-RADAR diffs \\
\end{tabularx}
\label{tab:rq2_results}

\end{table*}
Diff Risk Scoring is the main mechanism in RADAR that determines which diffs are eligible for more aggressive automation. A stricter threshold reduces risk but limits automation yield, while a relaxed threshold can widen the approval envelope and increase throughput. As part of calibration, we relaxed the Diff Risk Score threshold from p25 to p50.
Table~\ref{tab:rq2_results} shows that after calibration, the current RADAR-approve rate reached 60.31\% (L7) and the RADAR Verification pass rate is 26.31\% (L7). These results indicate that tuning the risk threshold improved the efficiency of the funnel, both by allowing more diffs to progress and by reducing friction at validation stages.
Safety is an explicit constraint for automated review/landing workflow. Across 535K+ RADAR-reviewed diffs, the revert rate is $\frac{1}{3}$ that of non-RADAR diffs, and the PI rate is $\frac{1}{50}$ that of non-RADAR diffs. While this is not a causal estimate, it is an important operational indicator that widening the approval envelope has not been associated with safety degradation at scale.
\begin{tcolorbox}
After relaxing the Diff Risk Score threshold from p25 to p50, the RADAR-approve rate reached 60.31\%. Across 535K+ RADAR-reviewed diffs, the revert rate is $\frac{1}{3}$ and the PI rate is $\frac{1}{50}$ that of non-RADAR diffs.
\end{tcolorbox}
\subsection{RQ3 (Impact): Review Efficiency}\label{sec:rq3_review_efficiency}

\textit{To what extent does automated review reduce end-to-end review latency for AI-generated code changes?}
\begin{table*}[t]

\centering
\caption{RQ3. Observed efficiency outcomes for diffs handled by RADAR.}

\begin{tabularx}{.90\textwidth}{X | r}
\textbf{Metric Name} & \textbf{Reported Value} \\ \hline
Median time to close (RADAR vs.\ human-reviewed diffs) & over 330\% reduction \\ \hline
Median diff review wall time (RADAR vs.\ human-reviewed diffs) & 35\% reduction \\ \hline
Scale of analysis & 535K+ diffs RADAR reviewed \\
\end{tabularx}
\label{tab:rq3_results}

\end{table*}
RADAR is intended to reduce time spent waiting for review and reduce review effort for eligible diffs, particularly for diffs generated by bots or AI prompts. This is important given the year over year growth in diff volume and diff size, and the observed decline in the percentage of diffs reviewed within 24 hours.
Table~\ref{tab:rq3_results} summarizes observed efficiency outcomes for diffs handled by RADAR. Across hundreds of thousands of RADAR-reviewed diffs, the median time to close is over 330\% lower and the median diff review wall time is 35\% lower than for human-reviewed diffs.
These results indicate that automation substantially reduces the time diffs spend waiting in review queues, which is consistent with RADAR removing the dependency on scarce human reviewer attention for eligible changes. In the context of rapidly increasing code creation, reducing review wall time is important because it can prevent low risk diffs from competing with higher risk diffs for reviewer attention.
\begin{tcolorbox}
Compared to human-reviewed diffs, RADAR reduces median time to close by over 330\% and median diff review wall time by 35\%, based on 535K+ diffs reviewed by RADAR.
\end{tcolorbox}
\section{Literature and Discussion}
\label{sec:literatureAndDiscussion}

In this section we position our contribution in the related work and discuss its importance.

\subsection{Modern Code Review}

Modern code review has evolved from formal software inspections into lightweight, tool-supported, asynchronous workflows~\cite{rigby2012contemporary,rigby2013convergent,bacchelli2013expectations}. Empirical studies show that review supports defect finding and process conformance~\cite{mcintosh2016emse,kononenko2015investigating,bosu2015characteristics}, and industrial case studies at Google and \Meta confirm these benefits at scale~\cite{sadowski2018modern,Shan2022FSE}. At the same time, review quality and participation are influenced by reviewer workload, seniority, and organizational factors~\cite{bosu2013impact,kononenko2015investigating}.
A persistent challenge is that review remains constrained by human attention and calendar time. Shan et al.\ used nudges to accelerate code review at \Meta~\cite{Shan2022FSE}, and reviewer recommendation systems aim to route changes to appropriate reviewers to reduce latency and improve quality~\cite{asthana2019whodo,balachandran2013reducing,ccetin2021review,Rigby2025TOSEM}. RADAR takes a complementary approach: rather than improving human review efficiency, it identifies diffs where human review can be safely replaced by automated validation, shifting scarce human attention toward higher risk changes.

\subsection{Risk-Aware Development and Defect Prediction}

Predicting the risk of software changes has a long history. Early work showed that change history features can predict fault-prone modules~\cite{GKMS99,MW00}, and subsequent research extended these models to just-in-time defect prediction at the commit level~\cite{kamei2016studying}. In industry, file-level risk scores have been used to prioritize refactoring and focus quality improvement effort~\cite{mockus2025codeimprovementpracticesmeta}.
RADAR builds on \Meta's Diff Risk Score (DRS), which applies risk prediction directly to the code review workflow. DRS powers approximately 20 risk-aware features across \Meta, including accept-to-ship, code freeze management, and risk-aware reviewer recommendation~\cite{Rigby2025TOSEM}. In RADAR, DRS serves as a gating mechanism that determines which diffs are eligible for automated review and landing, with configurable thresholds that allow each organization to calibrate its risk appetite. This represents a shift from using risk models for information, alerting reviewers to risky diffs, to using them for action, automating review decisions for low-risk diffs.

\subsection{AI-Assisted Coding and Coding Agents}

The emergence of large language models for code generation has transformed software development workflows. Code completion tools~\cite{chen2021evaluating,peng2023impact} have been widely adopted, and more recently, LLM-based coding agents that can autonomously write, test, and refactor code have appeared~\cite{wang2025agents,liu2024largelanguagemodelbasedagents,NEURIPS2024_5a7c9475}. Li et al.\ characterize this shift as SE 3.0, where autonomous coding agents are actively creating, reviewing, and evolving code at scale~\cite{li2025riseaiteammatessoftware}. At \Meta, code improvement activities including refactoring, dead code removal, and complexity reduction account for over 14\% of diffs~\cite{mockus2025codeimprovementpracticesmeta}, and RACER automates many of these tasks by generating diffs from predefined runbooks~\cite{shackleton2023dead}.
This rapid increase in AI-generated code creates direct pressure on review capacity. RADAR addresses the downstream consequence: as AI agents produce more diffs, the review pipeline must scale accordingly. RADAR's ACE pipeline is specifically designed to handle bot-authored diffs from systems like RACER and CodemodService, allowing low-risk automated changes to land without human review while routing higher risk changes to human reviewers.

\subsection{Automated Code Review}

Static analysis tools have long been integrated into code review workflows to catch bugs, style violations, and security issues before human reviewers examine the code~\cite{balachandran2013reducing,sadowski2018modern}. More recently, LLM-based approaches have been applied to automate parts of the review process itself. Industrial systems have explored using LLMs to generate review comments, detect code quality issues, and classify the risk level of changes.
RADAR's Automated Code Review (ACR) component uses an LLM to classify diff changes against safe and risk signal categories, making accept or reject decisions for the automated pipeline. Unlike general-purpose LLM review tools that aim to provide feedback to authors and reviewers, ACR is designed for a binary decision: whether a diff is safe enough to land without human review. This design reflects a specific operating point in the trade-off between automation coverage and safety, requiring high confidence (at least 8 out of 10) across all changes before approving a diff.

\subsection{Discussion}

Our findings address the three research questions in sequence.

\paragraph{Feasibility (RQ1).} RADAR demonstrates that an end-to-end production workflow for risk-aware automation of code review and landing is feasible at scale. The system has reviewed 535K+ diffs and landed 331K+ diffs, reaching 25K diffs/day, with a revert rate $\frac{1}{3}$ that of non-RADAR diffs and a PI rate $\frac{1}{50}$ that of non-RADAR diffs. Prior work on risk-aware review has focused on surfacing risk information to reviewers~\cite{Rigby2025TOSEM} or recommending better reviewers for risky diffs~\cite{asthana2019whodo}. RADAR inverts this approach by using risk scores to identify diffs that do not need human review, automating the entire review-to-landing pipeline for low-risk changes. An important part of this workflow is the differentiated eligibility model (Section~\ref{sec:diff_eligibility}), which applies distinct criteria to each automation source type: deterministic codemods bypass per-diff AI review entirely, AI-generated codemods require per-diff ACE evaluation, and RACER runbooks are governed by per-runbook risk history heuristics, daily volume caps, and configurable DRS thresholds. This source-specific eligibility design allows RADAR to calibrate the trade-off between automation throughput and safety at a granularity finer than the pipeline level. This operational record at scale complements prior benchmark-oriented evaluations of AI-assisted coding tools~\cite{NEURIPS2024_5a7c9475,li2025riseaiteammatessoftware} and extends the empirical understanding of how risk-aware automation performs in a large monorepo environment.
\paragraph{Calibration (RQ2).} Risk calibration is a practical operational lever. Widening the approval envelope from p25 to p50 was associated with higher automation yield and improved downstream validation, while safety outcomes remained stable. This finding is important because it suggests that initial conservative thresholds can be relaxed incrementally as operational confidence grows, and that the relationship between risk threshold and safety is not linear---a wider envelope can capture additional low-risk diffs without proportionally increasing incident rates.
\paragraph{Impact (RQ3).} Reducing time to close and review wall time for RADAR-reviewed diffs can mitigate review queue pressure and help shift scarce human attention toward higher-risk diffs. The observed reductions---over 330\% in median time to close and 35\% in median diff review wall time compared to human-reviewed diffs---suggest that automation removes a genuine bottleneck rather than merely accelerating already-fast diffs.
\paragraph{Knowledge transfer.} By analyzing the knowledge transfer implications of automated review, we highlight a dimension that has received limited attention in the automation literature: the trade-off between review efficiency and the knowledge diffusion that review provides~\cite{Rigby2014Peer}. Our network analysis shows that this trade-off is currently favorable, as RADAR-reviewed diffs are qualitatively different from the typical human-reviewed diff, but this balance may shift as automation coverage expands.
\section{Threats to Validity}\label{sec:threats}
\subsection{Generalizability}
This study was conducted at \Meta using RADAR and the surrounding review and deployment infrastructure. The organizational context includes a large monorepo, standardized tooling, and high automation coverage for testing and rollout. Other organizations may have different repository structures, review norms, risk tolerances, and validation infrastructure. These differences may affect both the feasibility of deploying a similar funnel and the magnitude of the observed throughput and efficiency changes.
Our results focus on low to medium complexity diffs that are eligible for RADAR automation and are evaluated using \Meta specific workflows, including eligibility policies, Diff Risk Scoring, LLM based automated code review, and deterministic validation checks. The selection criteria for eligible diffs may not match other codebases or other domains, especially where changes are larger, testing is less reliable, or release practices are less incremental. Finally, we cannot provide exact dates and do not release underlying diff level data due to privacy and confidentiality constraints, which may limit direct replication.
\subsection{Construct Validity}
Several outcomes are proxies. Time to close and diff review wall time capture responsiveness and queueing effects but do not directly measure defect prevention or long-term maintainability. Throughput metrics reflect scale but do not quantify the marginal value of replaced reviews.
Safety outcomes require careful interpretation. We report PI and revert rates as operational indicators, but they do not capture all forms of harm---regressions may be fixed without a PI, and some reverts may be precautionary. Incident attribution can also be imperfect when multiple changes land close in time.
The boundary between AI-generated and human-authored diffs is another construct threat, as AI assistance may be used in human-authored diffs, blurring category distinctions.
\subsection{Internal Validity}
Because diffs are not randomly assigned to RADAR, unobserved confounding can influence observed changes. Rollout expands gradually, eligibility policies evolve, and the composition of RADAR-eligible diffs may shift in complexity or test coverage over time. For RQ2, concurrent changes in onboarding, model behavior, or validation rules may coincide with the threshold change. For RQ3, difference-in-differences estimates depend on parallel-trend assumptions and stable measurement; we do not provide the full model specification. System-level interference is also possible: as RADAR absorbs low-risk volume, queue dynamics for remaining diffs may change, and some measured improvements may include these second-order effects.
\section{Conclusion}\label{sec:conclusion}
We presented RADAR, an end-to-end workflow that automates review and landing for eligible low-to-medium complexity diffs while routing higher-risk changes to human reviewers. Our evaluation, organized around feasibility, calibration, and impact, shows that risk-stratified automation can operate at production scale: RADAR has reviewed and landed hundreds of thousands of diffs with a revert rate $\frac{1}{3}$ and a PI rate $\frac{1}{50}$ that of non-RADAR diffs. Relaxing the risk threshold from p25 to p50 increased automation yield without degrading safety.  RADAR reduces median time to close by over 330\% and median diff review wall time by 35\% compared to human-reviewed diffs.
These findings suggest that layered automation---combining policy gates, risk scoring, LLM-based review, and deterministic validation---can absorb routine review volume created by AI-driven code growth, that risk thresholds are a practical calibration lever, and that the resulting latency reductions help shift scarce human attention toward changes where judgment is most beneficial. Future work should study longer-term effects on review backlogs, defect escape rates, and how eligibility policies can be tuned by code area and change type as coverage expands.
\bibliographystyle{ACM-Reference-Format}
\balance
\bibliography{main_nc}


\begin{thebibliography}{25}

%%% the Web address.

\ifx \showCODEN    \undefined \def \showCODEN     #1{\unskip}     \fi
\ifx \showDOI      \undefined \def \showDOI       #1{#1}\fi
\ifx \showISBNx    \undefined \def \showISBNx     #1{\unskip}     \fi
\ifx \showISBNxiii \undefined \def \showISBNxiii  #1{\unskip}     \fi
\ifx \showISSN     \undefined \def \showISSN      #1{\unskip}     \fi
\ifx \showLCCN     \undefined \def \showLCCN      #1{\unskip}     \fi
\ifx \shownote     \undefined \def \shownote      #1{#1}          \fi
\ifx \showarticletitle \undefined \def \showarticletitle #1{#1}   \fi
\ifx \showURL      \undefined \def \showURL       {\relax}        \fi
% invisible to TeX
\providecommand\bibfield[2]{#2}
\providecommand\bibinfo[2]{#2}
\providecommand\natexlab[1]{#1}
\providecommand\showeprint[2][]{arXiv:#2}

\bibitem[Asthana et~al\mbox{.}(2019)]%
        {asthana2019whodo}
\bibfield{author}{\bibinfo{person}{Sumit Asthana}, \bibinfo{person}{Rahul
  Kumar}, \bibinfo{person}{Ranjita Bhagwan}, \bibinfo{person}{Christian Bird},
  \bibinfo{person}{Chetan Bansal}, \bibinfo{person}{Chandra Maddila},
  \bibinfo{person}{Sonu Mehta}, {and} \bibinfo{person}{B Ashok}.}
  \bibinfo{year}{2019}\natexlab{}.
\newblock \showarticletitle{WhoDo: automating reviewer suggestions at scale}.
  In \bibinfo{booktitle}{\emph{Joint European Software Engineering Conference
  and Symposium on the Foundations of Software Engineering}}.
  \bibinfo{pages}{937--945}.
\newblock


\bibitem[Bacchelli and Bird(2013)]%
        {bacchelli2013expectations}
\bibfield{author}{\bibinfo{person}{Alberto Bacchelli} {and}
  \bibinfo{person}{Christian Bird}.} \bibinfo{year}{2013}\natexlab{}.
\newblock \showarticletitle{Expectations, outcomes, and challenges of modern
  code review}. In \bibinfo{booktitle}{\emph{International Conference on
  Software Engineering}}. \bibinfo{pages}{712--721}.
\newblock


\bibitem[Balachandran(2013)]%
        {balachandran2013reducing}
\bibfield{author}{\bibinfo{person}{Vipin Balachandran}.}
  \bibinfo{year}{2013}\natexlab{}.
\newblock \showarticletitle{Reducing human effort and improving quality in peer
  code reviews using automatic static analysis and reviewer recommendation}. In
  \bibinfo{booktitle}{\emph{International Conference on Software Engineering}}.
  \bibinfo{pages}{931--940}.
\newblock


\bibitem[Bosu and Carver(2013)]%
        {bosu2013impact}
\bibfield{author}{\bibinfo{person}{Amiangshu Bosu} {and}
  \bibinfo{person}{Jeffrey~C Carver}.} \bibinfo{year}{2013}\natexlab{}.
\newblock \showarticletitle{Impact of peer code review on peer impression
  formation: A survey}. In \bibinfo{booktitle}{\emph{International Symposium on
  Empirical Software Engineering and Measurement}}. \bibinfo{pages}{133--142}.
\newblock


\bibitem[Bosu et~al\mbox{.}(2015)]%
        {bosu2015characteristics}
\bibfield{author}{\bibinfo{person}{Amiangshu Bosu}, \bibinfo{person}{Michaela
  Greiler}, {and} \bibinfo{person}{Christian Bird}.}
  \bibinfo{year}{2015}\natexlab{}.
\newblock \showarticletitle{Characteristics of useful code reviews: An
  empirical study at microsoft}. In \bibinfo{booktitle}{\emph{International
  Working Conference on Mining Software Repositories}}.
  \bibinfo{pages}{146--156}.
\newblock


\bibitem[{\c{C}}etin et~al\mbox{.}(2021)]%
        {ccetin2021review}
\bibfield{author}{\bibinfo{person}{H~Alperen {\c{C}}etin},
  \bibinfo{person}{Emre Do{\u{g}}an}, {and} \bibinfo{person}{Eray
  T{\"u}z{\"u}n}.} \bibinfo{year}{2021}\natexlab{}.
\newblock \showarticletitle{A review of code reviewer recommendation studies:
  Challenges and future directions}.
\newblock \bibinfo{journal}{\emph{Science of Computer Programming}}
  (\bibinfo{year}{2021}), \bibinfo{pages}{102652}.
\newblock


\bibitem[Chen et~al\mbox{.}(2021)]%
        {chen2021evaluating}
\bibfield{author}{\bibinfo{person}{Mark Chen}, \bibinfo{person}{Jerry Tworek},
  \bibinfo{person}{Heewoo Jun}, \bibinfo{person}{Qiming Yuan},
  \bibinfo{person}{Henrique~Ponde de Oliveira~Pinto}, \bibinfo{person}{Jared
  Kaplan}, \bibinfo{person}{Harri Edwards}, \bibinfo{person}{Yuri Burda},
  \bibinfo{person}{Nicholas Joseph}, \bibinfo{person}{Greg Brockman},
  \bibinfo{person}{Alex Ray}, \bibinfo{person}{Raul Puri},
  \bibinfo{person}{Gretchen Krueger}, \bibinfo{person}{Michael Petrov},
  \bibinfo{person}{Heidy Khlaaf}, \bibinfo{person}{Girish Sastry},
  \bibinfo{person}{Pamela Mishkin}, \bibinfo{person}{Brooke Chan},
  \bibinfo{person}{Scott Gray}, \bibinfo{person}{Nick Ryder},
  \bibinfo{person}{Mikhail Pavlov}, \bibinfo{person}{Alethea Power},
  \bibinfo{person}{Lukasz Kaiser}, \bibinfo{person}{Mohammad Bavarian},
  \bibinfo{person}{Clemens Winter}, \bibinfo{person}{Philippe Tillet},
  \bibinfo{person}{Felipe~Petroski Such}, \bibinfo{person}{Dave Cummings},
  \bibinfo{person}{Matthias Plappert}, \bibinfo{person}{Fotios Chantzis},
  \bibinfo{person}{Elizabeth Barnes}, \bibinfo{person}{Ariel Herbert-Voss},
  \bibinfo{person}{William~Hebgen Guss}, \bibinfo{person}{Alex Nichol},
  \bibinfo{person}{Alex Paino}, \bibinfo{person}{Nikolas Tezak},
  \bibinfo{person}{Jie Tang}, \bibinfo{person}{Igor Babuschkin},
  \bibinfo{person}{Suchir Balaji}, \bibinfo{person}{Shantanu Jain},
  \bibinfo{person}{William Saunders}, \bibinfo{person}{Christopher Hesse},
  \bibinfo{person}{Andrew~N. Carr}, \bibinfo{person}{Jan Leike},
  \bibinfo{person}{Josh Achiam}, \bibinfo{person}{Vedant Misra},
  \bibinfo{person}{Evan Morikawa}, \bibinfo{person}{Alec Radford},
  \bibinfo{person}{Matthew Knight}, \bibinfo{person}{Miles Brundage},
  \bibinfo{person}{Mira Murati}, \bibinfo{person}{Katie Mayer},
  \bibinfo{person}{Peter Welinder}, \bibinfo{person}{Bob McGrew},
  \bibinfo{person}{Dario Amodei}, \bibinfo{person}{Sam McCandlish},
  \bibinfo{person}{Ilya Sutskever}, {and} \bibinfo{person}{Wojciech Zaremba}.}
  \bibinfo{year}{2021}\natexlab{}.
\newblock \bibinfo{title}{Evaluating Large Language Models Trained on Code}.
\newblock
\newblock
\showeprint[arxiv]{2107.03374}~[cs.LG]


\bibitem[Graves et~al\mbox{.}(2000)]%
        {GKMS99}
\bibfield{author}{\bibinfo{person}{T.~L. Graves}, \bibinfo{person}{A.~F. Karr},
  \bibinfo{person}{J.~S. Marron}, {and} \bibinfo{person}{H. Siy}.}
  \bibinfo{year}{2000}\natexlab{}.
\newblock \showarticletitle{Predicting fault incidence using software change
  history}.
\newblock \bibinfo{journal}{\emph{IEEE Transactions on Software Engineering}}
  \bibinfo{volume}{26}, \bibinfo{number}{2} (\bibinfo{year}{2000}).
\newblock


\bibitem[Kamei et~al\mbox{.}(2016)]%
        {kamei2016studying}
\bibfield{author}{\bibinfo{person}{Yasutaka Kamei}, \bibinfo{person}{Takafumi
  Fukushima}, \bibinfo{person}{Shane McIntosh}, \bibinfo{person}{Kazuhiro
  Yamashita}, \bibinfo{person}{Naoyasu Ubayashi}, {and}
  \bibinfo{person}{Ahmed~E Hassan}.} \bibinfo{year}{2016}\natexlab{}.
\newblock \showarticletitle{Studying just-in-time defect prediction using
  cross-project models}.
\newblock \bibinfo{journal}{\emph{Empirical Software Engineering}}
  \bibinfo{volume}{21}, \bibinfo{number}{5} (\bibinfo{year}{2016}),
  \bibinfo{pages}{2072--2106}.
\newblock


\bibitem[Kononenko et~al\mbox{.}(2015)]%
        {kononenko2015investigating}
\bibfield{author}{\bibinfo{person}{Oleksii Kononenko}, \bibinfo{person}{Olga
  Baysal}, \bibinfo{person}{Latifa Guerrouj}, \bibinfo{person}{Yaxin Cao},
  {and} \bibinfo{person}{Michael~W Godfrey}.} \bibinfo{year}{2015}\natexlab{}.
\newblock \showarticletitle{Investigating code review quality: Do people and
  participation matter?}. In \bibinfo{booktitle}{\emph{IEEE international
  conference on software maintenance and evolution (ICSME)}}.
  \bibinfo{pages}{111--120}.
\newblock


\bibitem[Li et~al\mbox{.}(2025)]%
        {li2025riseaiteammatessoftware}
\bibfield{author}{\bibinfo{person}{Hao Li}, \bibinfo{person}{Haoxiang Zhang},
  {and} \bibinfo{person}{Ahmed~E. Hassan}.} \bibinfo{year}{2025}\natexlab{}.
\newblock \bibinfo{title}{The Rise of AI Teammates in Software Engineering (SE)
  3.0: How Autonomous Coding Agents Are Reshaping Software Engineering}.
\newblock
\newblock
\showeprint[arxiv]{2507.15003}~[cs.SE]


\bibitem[Liu et~al\mbox{.}(2024)]%
        {liu2024largelanguagemodelbasedagents}
\bibfield{author}{\bibinfo{person}{Junwei Liu}, \bibinfo{person}{Kaixin Wang},
  \bibinfo{person}{Yixuan Chen}, \bibinfo{person}{Xin Peng},
  \bibinfo{person}{Zhenpeng Chen}, \bibinfo{person}{Lingming Zhang}, {and}
  \bibinfo{person}{Yiling Lou}.} \bibinfo{year}{2024}\natexlab{}.
\newblock \bibinfo{title}{Large Language Model-Based Agents for Software
  Engineering: A Survey}.
\newblock
\newblock
\showeprint[arxiv]{2409.02977}~[cs.SE]


\bibitem[McIntosh et~al\mbox{.}(2016)]%
        {mcintosh2016emse}
\bibfield{author}{\bibinfo{person}{Shane McIntosh}, \bibinfo{person}{Yasutaka
  Kamei}, \bibinfo{person}{Bram Adams}, {and} \bibinfo{person}{Ahmed~E.
  Hassan}.} \bibinfo{year}{2016}\natexlab{}.
\newblock \showarticletitle{{An Empirical Study of the Impact of Modern Code
  Review Practices on Software Quality}}.
\newblock \bibinfo{journal}{\emph{Empirical Software Engineering}}
  \bibinfo{volume}{21}, \bibinfo{number}{5} (\bibinfo{year}{2016}),
  \bibinfo{pages}{2146--2189}.
\newblock


\bibitem[Mockus et~al\mbox{.}(2025)]%
        {mockus2025codeimprovementpracticesmeta}
\bibfield{author}{\bibinfo{person}{Audris Mockus}, \bibinfo{person}{Peter~C
  Rigby}, \bibinfo{person}{Rui Abreu}, \bibinfo{person}{Anatoly Akkerman},
  \bibinfo{person}{Yogesh Bhootada}, \bibinfo{person}{Payal Bhuptani},
  \bibinfo{person}{Gurnit Ghardhora}, \bibinfo{person}{Lan~Hoang Dao},
  \bibinfo{person}{Chris Hawley}, \bibinfo{person}{Renzhi He},
  \bibinfo{person}{Sagar Krishnamoorthy}, \bibinfo{person}{Sergei Krauze},
  \bibinfo{person}{Jianmin Li}, \bibinfo{person}{Anton Lunov},
  \bibinfo{person}{Dragos Martac}, \bibinfo{person}{Francois Morin},
  \bibinfo{person}{Neil Mitchell}, \bibinfo{person}{Venus Montes},
  \bibinfo{person}{Maher Saba}, \bibinfo{person}{Matt Steiner},
  \bibinfo{person}{Andrea Valori}, \bibinfo{person}{Shanchao Wang}, {and}
  \bibinfo{person}{Nachiappan Nagappan}.} \bibinfo{year}{2025}\natexlab{}.
\newblock \showarticletitle{Metrics Driven Reengineering and Continuous Code
  Improvement at Meta}. In \bibinfo{booktitle}{\emph{International Conference
  on Automated Software Engineering}}.
\newblock


\bibitem[Mockus and Weiss(2000)]%
        {MW00}
\bibfield{author}{\bibinfo{person}{Audris Mockus} {and}
  \bibinfo{person}{David~M. Weiss}.} \bibinfo{year}{2000}\natexlab{}.
\newblock \showarticletitle{Predicting Risk of Software Changes}.
\newblock \bibinfo{journal}{\emph{Bell Labs Technical Journal}}
  \bibinfo{volume}{5}, \bibinfo{number}{2} (\bibinfo{date}{April--June}
  \bibinfo{year}{2000}), \bibinfo{pages}{169--180}.
\newblock


\bibitem[Peng et~al\mbox{.}(2023)]%
        {peng2023impact}
\bibfield{author}{\bibinfo{person}{Sida Peng}, \bibinfo{person}{Eirini
  Kalliamvakou}, \bibinfo{person}{Peter Cihon}, {and} \bibinfo{person}{Mert
  Demirer}.} \bibinfo{year}{2023}\natexlab{}.
\newblock \showarticletitle{The Impact of AI on Developer Productivity:
  Evidence from GitHub Copilot}. In \bibinfo{booktitle}{\emph{Arxiv}}.
\newblock
\urldef\tempurl%
\url{https://arxiv.org/abs/2302.06590}
\showURL{%
\tempurl}


\bibitem[Rigby et~al\mbox{.}(2012)]%
        {rigby2012contemporary}
\bibfield{author}{\bibinfo{person}{Peter Rigby}, \bibinfo{person}{Brendan
  Cleary}, \bibinfo{person}{Frederic Painchaud}, \bibinfo{person}{Margaret-Anne
  Storey}, {and} \bibinfo{person}{Daniel German}.}
  \bibinfo{year}{2012}\natexlab{}.
\newblock \showarticletitle{Contemporary peer review in action: Lessons from
  open source development}.
\newblock \bibinfo{journal}{\emph{IEEE software}}  \bibinfo{volume}{29}
  (\bibinfo{year}{2012}), \bibinfo{pages}{56--61}.
\newblock


\bibitem[Rigby and Bird(2013)]%
        {rigby2013convergent}
\bibfield{author}{\bibinfo{person}{Peter~C Rigby} {and}
  \bibinfo{person}{Christian Bird}.} \bibinfo{year}{2013}\natexlab{}.
\newblock \showarticletitle{Convergent contemporary software peer review
  practices}. In \bibinfo{booktitle}{\emph{International Symposium on the
  Foundations of Software Engineering}}. \bibinfo{pages}{202--212}.
\newblock


\bibitem[Rigby et~al\mbox{.}(2014)]%
        {Rigby2014Peer}
\bibfield{author}{\bibinfo{person}{Peter~C Rigby}, \bibinfo{person}{Daniel~M
  German}, \bibinfo{person}{Laura Cowen}, {and} \bibinfo{person}{Margaret-Anne
  Storey}.} \bibinfo{year}{2014}\natexlab{}.
\newblock \showarticletitle{Peer review on open-source software projects:
  Parameters, statistical models, and theory}.
\newblock \bibinfo{journal}{\emph{ACM Transactions on Software Engineering and
  Methodology (TOSEM)}} \bibinfo{volume}{23}, \bibinfo{number}{4}
  (\bibinfo{year}{2014}), \bibinfo{pages}{35}.
\newblock


\bibitem[Rigby et~al\mbox{.}(2025)]%
        {Rigby2025TOSEM}
\bibfield{author}{\bibinfo{person}{Peter~C. Rigby}, \bibinfo{person}{Seth
  Rogers}, \bibinfo{person}{Sadruddin Saleem}, \bibinfo{person}{Parth Suresh},
  \bibinfo{person}{Daniel Suskin}, \bibinfo{person}{Patrick Riggs},
  \bibinfo{person}{Chandra Maddila}, \bibinfo{person}{Nachiappan Nagappan},
  {and} \bibinfo{person}{Audris Mockus}.} \bibinfo{year}{2025}\natexlab{}.
\newblock \showarticletitle{Improving Code Reviewer Recommendation: Accuracy,
  Latency, Workload, and Bystanders}.
\newblock \bibinfo{journal}{\emph{ACM Trans. Softw. Eng. Methodol.}}
  (\bibinfo{date}{May} \bibinfo{year}{2025}).
\newblock
\showISSN{1049-331X}
\urldef\tempurl%
\url{https://doi.org/10.1145/3736405}
\showDOI{\tempurl}


\bibitem[Sadowski et~al\mbox{.}(2018)]%
        {sadowski2018modern}
\bibfield{author}{\bibinfo{person}{Caitlin Sadowski}, \bibinfo{person}{Emma
  S{\"o}derberg}, \bibinfo{person}{Luke Church}, \bibinfo{person}{Michal
  Sipko}, {and} \bibinfo{person}{Alberto Bacchelli}.}
  \bibinfo{year}{2018}\natexlab{}.
\newblock \showarticletitle{Modern code review: a case study at google}. In
  \bibinfo{booktitle}{\emph{Proceedings of the 40th International Conference on
  Software Engineering: Software Engineering in Practice}}. ACM,
  \bibinfo{pages}{181--190}.
\newblock


\bibitem[Shackleton et~al\mbox{.}(2023)]%
        {shackleton2023dead}
\bibfield{author}{\bibinfo{person}{Will Shackleton}, \bibinfo{person}{Katriel
  Cohn-Gordon}, \bibinfo{person}{Peter~C Rigby}, \bibinfo{person}{Rui Abreu},
  \bibinfo{person}{James Gill}, \bibinfo{person}{Nachiappan Nagappan},
  \bibinfo{person}{Karim Nakad}, \bibinfo{person}{Ioannis Papagiannis},
  \bibinfo{person}{Luke Petre}, \bibinfo{person}{Giorgi Megreli},
  {et~al\mbox{.}}} \bibinfo{year}{2023}\natexlab{}.
\newblock \showarticletitle{Dead Code Removal at Meta: Automatically Deleting
  Millions of Lines of Code and Petabytes of Deprecated Data}. In
  \bibinfo{booktitle}{\emph{Proceedings of the 31st ACM Joint European Software
  Engineering Conference and Symposium on the Foundations of Software
  Engineering}}. \bibinfo{pages}{1705--1715}.
\newblock


\bibitem[Shan et~al\mbox{.}(2022)]%
        {Shan2022FSE}
\bibfield{author}{\bibinfo{person}{Qianhua Shan}, \bibinfo{person}{David
  Sukhdeo}, \bibinfo{person}{Qianying Huang}, \bibinfo{person}{Seth Rogers},
  \bibinfo{person}{Lawrence Chen}, \bibinfo{person}{Elise Paradis},
  \bibinfo{person}{Peter~C. Rigby}, {and} \bibinfo{person}{Nachiappan
  Nagappan}.} \bibinfo{year}{2022}\natexlab{}.
\newblock \showarticletitle{Using nudges to accelerate code reviews at scale}.
  In \bibinfo{booktitle}{\emph{Proceedings of the 30th ACM Joint European
  Software Engineering Conference and Symposium on the Foundations of Software
  Engineering}} \emph{(\bibinfo{series}{ESEC/FSE 2022})}.
  \bibinfo{publisher}{Association for Computing Machinery},
  \bibinfo{pages}{472--482}.
\newblock
\urldef\tempurl%
\url{https://doi.org/10.1145/3540250.3549104}
\showDOI{\tempurl}


\bibitem[Wang et~al\mbox{.}(2025)]%
        {wang2025agents}
\bibfield{author}{\bibinfo{person}{Yanlin Wang}, \bibinfo{person}{Wanjun
  Zhong}, \bibinfo{person}{Yanxian Huang}, \bibinfo{person}{Ensheng Shi},
  \bibinfo{person}{Min Yang}, \bibinfo{person}{Jiachi Chen},
  \bibinfo{person}{Hui Li}, \bibinfo{person}{Yuchi Ma},
  \bibinfo{person}{Qianxiang Wang}, {and} \bibinfo{person}{Zibin Zheng}.}
  \bibinfo{year}{2025}\natexlab{}.
\newblock \showarticletitle{Agents in software engineering: Survey, landscape,
  and vision}.
\newblock \bibinfo{journal}{\emph{Automated Software Engineering}}
  \bibinfo{volume}{32}, \bibinfo{number}{2} (\bibinfo{year}{2025}),
  \bibinfo{pages}{1--36}.
\newblock


\bibitem[Yang et~al\mbox{.}(2024)]%
        {NEURIPS2024_5a7c9475}
\bibfield{author}{\bibinfo{person}{John Yang}, \bibinfo{person}{Carlos~E.
  Jimenez}, \bibinfo{person}{Alexander Wettig}, \bibinfo{person}{Kilian
  Lieret}, \bibinfo{person}{Shunyu Yao}, \bibinfo{person}{Karthik Narasimhan},
  {and} \bibinfo{person}{Ofir Press}.} \bibinfo{year}{2024}\natexlab{}.
\newblock \showarticletitle{SWE-agent: Agent-Computer Interfaces Enable
  Automated Software Engineering}. In \bibinfo{booktitle}{\emph{Advances in
  Neural Information Processing Systems}}, Vol.~\bibinfo{volume}{37}.
  \bibinfo{publisher}{Curran Associates, Inc.}, \bibinfo{pages}{50528--50652}.
\newblock


\end{thebibliography}
\end{document}